\documentstyle[12pt,epsfig,amsmath,amssymb,pstricks]{article}
\begin{document}

\title{\bf Implementation of classical communication in a quantum world}

\author{{Chris Fields}\\ \\
{\it 815 East Palace \# 14}\\
{\it Santa Fe, NM 87501 USA}\\ \\
{fieldsres@gmail.com}}
\maketitle

\begin{abstract}
Observations of quantum systems carried out by finite observers who subsequently communicate their results using classical data structures can be described as ``local operations, classical communication'' (LOCC) observations.  The implementation of LOCC observations by the Hamiltonian dynamics prescribed by minimal quantum mechanics is investigated.  It is shown that LOCC observations cannot be described using decoherence considerations alone, but rather require the \textit{a priori} stipulation of a positive operator-valued measure (POVM) about which communicating observers agree.  It is also shown that the transfer of classical information from system to observer can be described in terms of system-observer entanglement, raising the possibility that an apparatus implementing an appropriate POVM can reveal the entangled system-observer states that implement LOCC observations.
\end{abstract}


\section{Introduction}

Suppose spatially-separated observers Alice and Bob each perform local measurements on a spatially-extended quantum system - for example, a pair of entangled qubits in an asymmetric Bell state - and afterwards communicate their experimental outcomes to each other.  This ``local operations, classical communication'' (LOCC, e.g. \cite{nielsen-chaung:00} Ch. 12) scenario characterizes quantum key distribution, preparation of the initial states and subsequent observation of the final states of quantum computers, and practical laboratory investigations of spatially-extended quantum systems; indeed LOCC characterizes all situations in which two or more observers interact with a quantum system and then report their observations by encoding them into sharable classical data structures.  Formal descriptions of LOCC scenarios generally specify the quantum system $\mathbf{S}$ with which the observers interact by explicitly specifying its quantum degrees of freedom and hence its Hilbert space $\mathcal{H}_{\mathbf{S}}$; in addition, they typically explicitly specify the ``prepared'' quantum state $|\mathbf{S}\rangle$ with which the observers interact, for example by an expression such as `$|\mathbf{S}\rangle = (1/ \surd 2)(|0\rangle_{\mathbf{A}} \otimes |1\rangle_{\mathbf{B}} - |1\rangle_{\mathbf{A}} \otimes |0\rangle_{\mathbf{B}})$' where $|0\rangle$ and $|1\rangle$ are basis vectors and `$\mathbf{A}$' and `$\mathbf{B}$' name Alice and Bob, respectively.  The ``local operations'' are generally dealt with cursorily: Alice and Bob are said to measure spin or polarization, for example, with the details of the apparatus used to do so, if any are given, relegated to the Methods section.  The ``classical communication'' between Alice and Bob is rarely discussed at all.  Understanding LOCC in physical terms, however, requires not just understanding the quantum state being observed, but understanding both the ``local operations'' and the ``classical communication'' as physical processes.

Let us begin with classical communication.  Any finite message from Bob to Alice can be represented as a finite sequence of classical bits.  It must, moreover, be encoded in some physical medium \cite{landauer:99} - notes in a logbook, for example, or an email message, or coherent vibrations of air molecules.  Bob encodes the message and Alice receives it by performing local operations on the physical medium employed for transmission.  Successful transmission requires, therefore, that Alice monitor the medium for messages, and that Alice and Bob share an encoding/decoding scheme - a data structure with \texttt{write} and \texttt{read} methods - as well as a semantics for that data structure that renders the message meaningful.  These requirements are independent of whether Alice and Bob are human beings or non-human information-processing machines; two computers attached to the internet must share a communication protocol (e.g. \texttt{tcp/ip}) and must share assumptions about both the syntax and semantics of the data structures employed to encode transmitted messages.

The local operations performed by Alice and Bob have, therefore, two distinct targets.  Alice and Bob must each operate locally on $\mathbf{S}$ to extract classical information, and must each operate locally on their shared communication medium to either encode (Bob) or decode (Alice, and Bob if he checks his encoding) the classical information contained in the transmitted message.  Most discussions of LOCC acknowledge that the interactions with $\mathbf{S}$ involve quantum measurement; most neglect the fact that, if quantum theory is assumed to be universal, the encoding and decoding steps also involve interactions with a quantum system: the physical medium of communication.  Most, moreover, neglect the fact that Alice and Bob are themselves quantum systems.  The purpose of the present paper is to examine LOCC from a perspective that acknowledges these facts; it is, therefore, to ask what is required to implement LOCC in a quantum world.

The next section, ``Preliminaries'' discusses the fundamental assumption that quantum theory is universal and two of its consequences: that the extraction of classical information from quantum systems can be represented by the action of positive operator-valued measures (POVMs, reviewed by \cite{nielsen-chaung:00} Ch. 2), and that observers must deploy POVMs to \textit{identify} quantum systems of interest.  The third section, ``Decompositional equivalence and its consequences'' discusses a second fundamental assumption: that the universe as a whole exhibits a symmetry, \textit{decompositional equivalence}, that allows alternative tensor-product structures (TPSs) for a single Hilbert space \cite{fields:12a}.  Like the assumption of universality, decompositional equivalence is an empirical assumption; if it is true, physical dynamics cannot depend in any way on TPSs that may be specified as defining ``systems'' of interest.  In a universe satisfying decompositional equivalence, system-environment decoherence, which depends for its definition on the specification of a TPS, can have no \textit{physical} consequences, and hence can neither create nor alter physical encodings of classical information.   Observers cannot, therefore, take for granted physical encodings by their shared environment of either the boundaries or the pointer states of specific systems of interest, as is proposed in the ``environment as witness'' formulation of decoherence theory \cite{zurek:04, zurek:05} and quantum Darwinism \cite{zurek:06, zurek:09}.  The fourth section, ``Decoherence as semantics'' shows that decoherence can be represented as the action of a POVM, and hence as being a semantic or model-theoretic mapping from physical systems to classical data structures, and in particular to classical virtual machines.  It is shown that the semantic consistency conditions for constructing such mappings are those familiar from the consistent histories formulation of quantum measurement (e.g. \cite{griffiths:02}).  The fifth section, ``Observation as entanglement'' returns to the question of how multiple observers in a LOCC setting identify and determine the state of a single system and then communicate their results.  It shows that LOCC requires an infinite regress of assumptions regarding prior classical communications between the observers involved.  In the absence of further assumptions, therefore, observations under LOCC conditions cannot be carried out in a universe characterized by both universal quantum theory and decompositional equivalence.  It then shows that the classical correlation between the states of an observer and an observed system produced by the action of a POVM would result from observer-system entanglement, and that such a correlation would be perfect if the entanglement was monogamous.  Hence observation mediated by a POVM can be regarded as alternative formal description of quantum entanglement; the transfer of classical information such entanglement enables is independent of system boundaries and relative, for any third party, to the specification of an appropriate basis for the joint system-observer state.  While this result renders the explanation of classical communication in terms of an observer-independent physical process of ``emergence'' unattainable, it offers the possibility that an apparatus implementing an appropriate POVM could reveal the specific system-observer entanglements that implement the observation of classical outcomes.  The paper concludes that the appearance of shared, public classicality in the physical world is fully analogous to the appearance of algorithm instantiation in classical computer science: both are cases of a shared, jointly stipulated semantic interpretation.

\section{Preliminaries}

\subsection{Assumption: Quantum theory is universal}

The first and most fundamental assumption made here is that quantum theory is universal: all \textit{physical} systems are \textit{quantum} systems.  The universe $\mathbf{U}$, in particular, is a physical system; it is therefore a quantum system, and can be characterized by a Hilbert space $\mathcal{H}_{\mathbf{U}}$ comprising a collection of quantum degrees of freedom.  The universe is moreover, as assumed by Everett \cite{everett:57}, not part of anything else; it is an \textit{isolated} quantum system.  The evolution of the universal quantum state $|\mathbf{U}\rangle$, therefore, satisfies a Schr\"odinger equation $(\partial / \partial t) | \mathbf{U} \rangle = - (\imath / \hbar) \mathit{H}_{\mathbf{U}} | \mathbf{U} \rangle$, where $H_{\mathbf{U}}$ is a deterministic universal Hamiltonian.  This assumption rules out any objective non-unitary ``collapse'' of $| \mathbf{U} \rangle$; it amounts to the adoption of what Landsman \cite{landsman:07} calls ``stance 1'' regarding quantum theory, a stance that is realist about quantum states, and therefore demands an explanation for the appearance of classicality.  All available experimental evidence is consistent with this universality assumption \cite{schloss:06}.  Alice, Bob, the systems that they observe and the systems that they employ to encode classical communications are, on this assumption, all collections of quantum degrees of freedom evolving under the action of the universal Hamiltonian $H_{\mathbf{U}}$. 

The assumption that all physical systems are quantum systems clearly does not entail that all \textit{descriptions} of physical systems are quantum-theoretical descriptions.  Some descriptions are classical; others are quantum-theoretical.  Classical descriptions of physical systems are in some cases (e.g. for billiard balls) sufficient for practical purposes, while in other cases (e.g. for electrons) they are not.  The observable world \textit{appears} classical to human observers employing their unaided senses; this appearance will be referred to as ``observational classicality.''  Human observers, moreover, record and communicate their observations using classical data structures, as do all artificial observers thus far constructed by humans.  Hence all descriptions of physical, i.e. quantum systems, whether they are classical descriptions or quantum-theoretical descriptions, are both recorded for future access and communicated using classical data structures, regardless of whether the observers involved are humans or artifacts.  It is this classicality of recorded descriptions that both motivates and requires LOCC as a characterization of the interaction of multiple observers with a quantum, i.e. physical system.

Under the assumption of universality, understanding the requirements of LOCC in the case of either Alice or Bob individually clearly requires understanding quantum measurement, and in particular understanding whether observational classicality can be supposed to ``emerge'' from the dynamics specified by $H_{\mathbf{U}}$.  If the observed system $\mathbf{S}$ is regarded as a quantum information processor, this question of observational classicality becomes the question of how the behavior of $\mathbf{S}$ can be interpreted as computation.  How, for example, do the unitary transformations of the quantum state of a quantum Turing machine (QTM, \cite{deutsch:85}) or Hamiltonian oracle \cite{farhi:96} implement a computation on a classical data structure encoded by the system's initial state?  In what sense do the events that occur between measurements in a measurement-based quantum computer \cite{briegel:09} implement computation?  That these questions are both foundational to quantum computing and non-trivial has been emphasized by Aaronson \cite{aaronson:05}.

What the LOCC concept adds to the quantum measurement problem as traditionally presented (e.g. \cite{landsman:07, wallace:08}) is the requirement that two observers interact with the \textit{same} system, and then moreover interact, via a communicated message, with each other.  Understanding LOCC, therefore, requires understanding measurement as both a \textit{redundant} or \textit{repeatable} process and as a \textit{social} process; with the exception of some discussions of Wigner's friend, neither of these aspects of LOCC are considered in traditional accounts of single-observer measurement.  It will be shown below, in \S 3-4 and \S 5 respectively, that the theoretical issues raised by these additional considerations are non-trivial.

\subsection{Consequence: Measurements are actions by POVMs}

If quantum theory is universal, measurements can be represented by POVMs.  A POVM is a collection $\lbrace E_{i} \rbrace$ of positive-semidefinite Hilbert-space automorphisms that have been normalized so as to sum to unity; POVMs generalize traditional projective measurements (e.g. \cite{vonNeumann:32}) by dropping the requirement of orthogonality and hence the requirement that all elements of a measurement project onto the same Hilbert-space basis.  If $\lbrace E_{i}^{\mathbf{S}} \rbrace$ is a POVM representing a measurement of the state of some quantum system $\mathbf{S}$, then each component $E_{j}^{\mathbf{S}} \in \lbrace \mathit{E_{i}}^{\mathbf{S}} \rbrace$ is a Hilbert-space automorphism on $\mathcal{H}_{\mathbf{S}}$, i.e. $E_{j}^{\mathbf{S}}: \mathcal{H}_{\mathbf{S}} \rightarrow \mathcal{H}_{\mathbf{S}}$; one can also write $E_{j}^{\mathbf{S}}: |\mathbf{S}\rangle \mapsto |\mathbf{S}^{\prime}\rangle$, where in general $|\mathbf{S}\rangle \neq |\mathbf{S}^{\prime}\rangle$.  Given the assumption of universality, it is clear that any such automorphism must be \textit{implemented} by the unitary physical propagator $e^{-(\imath / \hbar) \mathit{H}_{\mathbf{U}}\mathit{t}}$ acting on the universal Hilbert space $\mathcal{H}_{\mathbf{U}}$, and hence on $|\mathbf{S}\rangle$ as a collection of components of some universal state $|\mathbf{U}\rangle$.  Hence a measurement can be thought of as a physical \textit{action} by a POVM, as emphasized for example by Fuchs' \cite{fuchs:10} depiction of a POVM as an observer's prosthetic hand.

Treating a POVM as a collection of Hilbert-space automorphisms does not, however, capture the sense in which observations \textit{extract} classical information from quantum systems.  To see how POVMs model measurement, it is useful to return to the case of a POVM with mutually orthognal components, i.e. a von Neumann projection $\lbrace \Pi_{i} \rbrace$ defined on a Hilbert space $\mathcal{H}$.  Each component $\Pi_{j}$ of a von Neumann projection $\lbrace \Pi_{i} \rbrace$ projects any state $|\psi\rangle \in \mathcal{H}$ onto a basis vector $|j\rangle$ of $\mathcal{H}$.  If the set $\lbrace |i\rangle \rbrace$ of images of the components of $\lbrace \Pi_{i} \rbrace$ is complete in the sense of spanning $\mathcal{H}$, one can write $|\psi\rangle = \sum_{j} \alpha_{j} |j\rangle$ for states $|\psi\rangle \in \mathcal{H}$.  In this case a general Hermitian observable $M$ can be written $M = \sum_{j} \alpha_{j} \Pi_{j}$ where $\alpha_{j}$ is the $j^{th}$ possible observable outcome of $M$ acting on $|\psi\rangle$.  Hence from an observer's point of view, what a projection $\lbrace \Pi_{i} \rbrace$ produces is not just a new state vector, but a \textit{real outcome value} $\alpha_{j}$; $\lbrace \Pi_{i} \rbrace$ is not just a Hilbert-space automorphism, but is also a mapping from $\mathcal{H}$ to the set of real outcome values of some observable of interest.

A general POVM $\lbrace E_{i}^{\mathbf{S}} \rbrace$ can be thought of as a mapping from $\mathcal{H}_{\mathbf{S}}$ to a set of real outcome values with two caveats.  First, the components of a general POVM are not necessarily orthogonal and hence do not, in general, all project to the same basis.  Second, any finite observer can explicitly represent, and hence can physically encode in a classical memory or communication medium, values with at most some finite number $N$ of bits.  Hence from an observer's point of view, a component $E_{j}^{\mathbf{S}}$ of a general POVM $\lbrace E_{i}^{\mathbf{S}} \rbrace$ is not just an automorphism on $\mathcal{H}_{\mathbf{S}}$; it is also a mapping $E_{j}^{\mathbf{S}}: \mathcal{H}_{\mathbf{S}} \rightarrow (\mathbb{B}^{\mathit{N}}, \{\mathbf{basis}\}^{\mathbf{S}})$, where $\mathbb{B}^{\mathit{N}}$ is the set of binary codes of length $N$ and $\{\mathbf{basis}\}^{\mathbf{S}}$ is the set of bases of $\mathcal{H}_{\mathbf{S}}$ \cite{fields:12a}.  Indeed, any collection $\lbrace F_{i}^{\mathbf{S}} \rbrace$ of mappings $F_{j}^{\mathbf{S}}: \mathcal{H}_{\mathbf{S}} \rightarrow (\mathbb{B}^{\mathit{N}}, \{\mathbf{basis}\}^{\mathbf{S}})$ for which the probabilities $P(\alpha_{j})$ of obtaining real outcome values $\alpha_{j} \in \mathbb{B}^{\mathit{N}}$ sum to unity, and for which each of the components $F_{j}^{\mathbf{S}}$ is implementable by the unitary physical propagator $e^{-(\imath / \hbar) \mathit{H}_{\mathbf{U}}\mathit{t}}$ acting on the universal Hilbert space $\mathcal{H}_{\mathbf{U}}$ must be positive semi-definite (to yield real outcome values), normalized (to yield well-defined probabilities) and be a collection of Hilbert-space automorphisms (to be implementable by $e^{-(\imath / \hbar) \mathit{H}_{\mathbf{U}}\mathit{t}}$); hence such a collection must be a POVM.  The POVM formalism thus represents the extraction of classical information from quantum systems in the only way that it can be represented while maintaining consistency with the universality assumption.

The assumption that all measurements can be represented by POVMs clearly does not entail that an observer can explicitly write down the components of every POVM that he or she might deploy in the course of interacting with the world.  Doing so in any particular case would require both a complete specification of the outcome values obtainable with that POVM and a complete specification of the Hilbert space upon which it acts, or as discussed below, a complete specification of the inverse image in $\mathcal{H}_{\mathbf{U}}$ of its set of obtainable outcome values.  Such a specification would, for any particular POVM and hence any particular Hilbert space, require scientific investigation of the physical system represented by that Hilbert space to be \textit{complete}.  Classical theorems \cite{ashby:56, moore:56} restricting the completeness of system identification strongly suggest that such completeness is infeasible in principle.  Hence explicitly-specified POVMs can at best be viewed as predictively-adequate approximations based on experimental investigations carried out thus far; in practice such POVMs are available only for systems with small numbers of (known or stipulated) degrees of freedom.

\subsection{Consequence: Observers must identify the systems they observe}

When a new graduate student enters a laboratory, he or she is introduced to the various items of apparatus that the laboratory employs.  The reason for this ritual is obvious: the student cannot be expected to reliably report the state of a particular apparatus if he or she cannot \textit{identify} that apparatus.  Traditional discussions of quantum measurement take the ability of observers to identify items of apparatus for granted.  For example, Ollivier, Poulan and Zurek define ``objectivity'' for physical systems operationally as follows:

\begin{quote}
``A property of a physical system is \textit{objective} when it is:
\begin{list}{\leftmargin=2em}
\item
1. simultaneously accessible to many observers,
\item
2. who are able to find out what it is without prior knowledge about the system of interest, and 
\item
3. who can arrive at a consensus about it without prior agreement.''
\end{list}
\begin{flushright}
(p. 1 of \cite{zurek:04}; p. 3 of \cite{zurek:05})
\end{flushright}
\end{quote}
Nothing is said in this definition, or in the surrounding discussion \cite{zurek:04, zurek:05}, about how observers are able to ``access'' a physical system ``without prior knowledge'' of such state variables as its location, size or shape, and without ``prior agreement'' about which item in their shared environment constitutes the system of interest.  To find the identification of physical systems by observers treated explicitly, one must look to cybernetics, where unique identification of even classical finite-state machines (FSMs) by finite sequences of finite observations is shown to be impossible in principle \cite{ashby:56, moore:56}, or to the cognitive neuroscience of perception, where the identification in practice of individual systems over extended periods of time is recognized as a computationally-intensive heuristic process \cite{rips:06, scholl:07, fields:12b}.

In practice, observers identify items of laboratory apparatus by finite sets of classically-specified criteria: location, size, shape, color, overall appearance, laboratory-affixed labels, brand name.  These criteria are encodable as finite binary strings.  If quantum theory is universal, items of laboratory apparatus are quantum systems, and hence are characterizable by Hilbert spaces comprising their quantum degrees of freedom.  Observing a laboratory apparatus, therefore, requires deploying an operator that maps a collection of quantum degrees of freedom to a finite set of finite binary strings; by the reasoning above, such operators can only be POVMs.  \textit{Identifying} a system of interest clearly requires \textit{observing} it; hence an observer can only identify a system of interest by deploying a POVM.  Call POVMs deployed to identify systems of interest ``system-identifying'' POVMs.  For simplicity, a system-identifying POVM can be regarded as yielding as output just the conventionalized name of the system it identifies, e.g. `$\mathbf{S}$' or `the Canberra\textsuperscript{\textregistered} Ge(Li) detector' \cite{fields:12a}.

The formal definition of system-identifying POVMs is complicated by two related issues.  First, the vast majority of systems identified by human observers are characterized, like laboratory apparatus are characterized, not by possible outcome values of their quantum degrees of freedom, but by possible outcome values of bulk degrees of freedom such as macroscopic size or shape.  The exceptions - the systems that those who reject the universality of quantum theory consider to be the only \textit{bona fide} ``quantum systems'' - are systems \textit{defined by} particular values of quantum degrees of freedom, as electrons or the Higgs boson are currently defined within the Standard Model, or are systems \textit{defined by} certain observable behaviors of macroscopic apparatus, as electrons were defined in the late $19^{th}$ century.  The second complication is that observers, as emphasized by Zurek \cite{zurek:98, zurek:03} and others, typically interact not with systems of interest themselves, but with their surrounding environments.  While in the case of macroscopic systems such as laboratory apparatus this environment may be treated using a straightforward approximation, for example as the ambient photon field, in the case of either microscopic or very distant systems it is complicated by the inclusion of laboratory apparatus; our interactions with presumptive Higgs bosons, for example, are via an environment containing the ATLAS \cite{atlas:12} or CMS \cite{cms:12} detectors.  These complicating issues are not significantly simplified by considering non-human observers; the components of such observers that record classical records are, with the exception of such things as blocks of plastic that record the passage of cosmic rays, almost as distant from the microscopic events to which their records refer as are their human minders.

In recognition of the role of the intervening environment in the observation and hence identification of systems of interest, it has been proposed that system-identifying POVMs be defined, in general, over either the physically-implemented information channel with which an observer interacts (i.e. the observer's environment) \cite{fields:12a} or over the universe $\mathbf{U}$ as a whole \cite{fields:12c}.  The latter definition is adopted here, as it simplifies the description of LOCC by allowing two or more observers to be regarded as deploying the \textit{same} system-identifying POVM.  Defining system-identifying POVMs over all of $\mathbf{U}$ acknowledges, moreover, the actual epistemic position of any finite observer. Observations are information-transferring actions by the observer's environment on the observer.  Without a complete, deterministic theory of the behavior of $\mathbf{U}$, such actions cannot be predicted precisely; without sufficient recording capacity to record the state of every degree of freedom of $\mathbf{U}$ at the instant of observation, such actions cannot be replicated precisely.  Any finite observer can, therefore, at best predict or retrodict only approximately and heuristically what degrees of freedom of $\mathbf{U}$ might be causally responsible for any particular episode of observation.  An observer \textit{can}, however, be sure that such degrees of freedom are within $\mathbf{U}$, so defining system-identifying POVMs over $\mathbf{U}$ can be viewed as an exercise of epistemic conservatism.

Defining system-identifying POVMs over $\mathbf{U}$ as a whole does not render observations nonlocal.  Any finite observer must expend finite energy to record the outcomes obtained by deploying a POVM; hence any observation requires finite time.  Any finite observer can, moreover, deploy a POVM for only a finite time.  A finite observer can, therefore, regard a system-identifying POVM - or any POVM - as extracting classical information from at most a local volume with a horizon at $c \Delta t$, where $\Delta t$ is the period of observation.  \textit{Quantum} information may originate outside this volume by entanglement, but such entanglement is undetectable in principle by the observer.  Alice can only regard classical information extracted from a quantum system employed as a communication channel as a message from Bob if Bob is in her light-cone; LOCC requires timelike, not spacelike, separation of observers.

Defining system-identifying POVMs over $\mathbf{U}$ as a whole does not, moreover, resolve the question of how such POVMs - or how any POVMs - can yield outcome values for bulk degrees of freedom such as macroscopic size or shape.  This question is, clearly, the question of quantum measurement itself; in particular, it is the question of the ``emergence of classicality'' that is taken up in \S 4 below.

\section{Decompositional equivalence and its consequences}

\subsection{Assumption: Our universe exhibits decompositional equivalence}

A fundamental requirement of observational objectivity, and hence of science as practiced, is that reality is independent of the language chosen to describe it.  This fundamental assumption that reality is independent of the descriptive terms and hence the semantics chosen by observers - in particular, human observers - underlies the assumption in scientific practice that any arbitrary collection of physical degrees of freedom can be stipulated to be a ``system of interest'' and named with a symbol such as `$\mathbf{S}$' without this choice of language affecting either fundamental physical laws or their outcomes as expressed by the dynamical behavior of the degrees of freedom contained within $\mathbf{S}$.  It similarly underlies the assumption that, given the technological means, an experimental apparatus to investigate the behavior of $\mathbf{S}$ can be designed and constructed without altering either fundamental physical laws or the dynamical behavior of the degrees of freedom contained within $\mathbf{S}$.  These assumptions operate prior to apparatus-dependent experimental interventions into the behavior of $\mathbf{S}$, and hence prior to \textit{observations} of $\mathbf{S}$, both logically and, in the course of practical investigations of microscopic degrees of freedom by means of macroscopic apparatus, temporally.

This fundamental assumption that reality is independent of semantics can be generalized to state an assumed dynamical symmetry: the universal dynamics $H_{\mathbf{U}}$ is asumed to be independent of, and hence symmetric under arbitrary modifications of, boundaries drawn in $\mathcal{H}_{\mathbf{U}}$ by  specifications of tensor product structures.  Call this symmetry \textit{decompositional equivalence} \cite{fields:12a}.  Stated formally, decompositional equivalence is the assumption that if $\mathbf{S \otimes E} = \mathbf{S^{\prime} \otimes E^{\prime}} = \mathbf{U}$, the dynamics $H_{\mathbf{U}} = H_{\mathbf{S}} + H_{\mathbf{E}} + H_{\mathbf{S-E}} = H_{\mathbf{S^{\prime}}} + H_{\mathbf{E^{\prime}}} + H_{\mathbf{S^{\prime}-E^{\prime}}}$, where $\mathbf{S}$ and $\mathbf{S^{\prime}}$ are arbitrarily chosen collections of physical degrees of freedom, $\mathbf{E}$ and $\mathbf{E^{\prime}}$ are their respective ``environments'' and $H_{\mathbf{S-E}}$ and $H_{\mathbf{S^{\prime}-E^{\prime}}}$ are, respectively, the $\mathbf{S-E}$ and $\mathbf{S^{\prime}-E^{\prime}}$ interaction Hamiltonians.  Such equivalence of TPSs of $\mathcal{H}_{\mathbf{U}}$ can be alternatively expressed in terms of the linearity of $H_{\mathbf{U}}$: if $H_{\mathbf{U}} = \sum_{ij} H_{ij}$ where the indices $i$ and $j$ range without restriction over all quantum degrees of freedom within $\mathcal{H}_{\mathbf{U}}$, decompositional equivalence is the assumption that the interaction matrix elements $\langle i|H_{ij}|j\rangle$  do not depend on the labels assigned to collections of degrees of freedom by specifications of TPSs.  Decompositional equivalence is thus consistent with the general philosophical position of microphysicalism (for a recent review, see \cite{hu:12}), but involves no claims about explanatory reduction, and indeed no claims about explanation at all; it requires only that emergent properties of composite objects exactly supervene, as a matter of physical fact, on the fundamental interactions of the microscale components of those objects.

As is the assumption that quantum theory is universal, the assumption that the universe satisfies decompositional equivalence is an empirical assumption.  Its empirical content is most obvious in its formulation as the assumption that interaction matrix elements $\langle i|H_{ij}|j\rangle$  do not depend on specifications of TPSs.  This is an assumption that the pairwise interaction Hamiltonians $H_{ij}$ are not just independent of where and when the degrees of freedom labeled by $i$ and $j$ interact, but are also independent of any other classical information that might be included in the specification of a reference frame from which the interaction of $i$ and $j$ might be observed.  As such, it is similar in spirit to Tegmark's ``External Reality Hypothesis (ERH)'' that ``there exists an external physical reality completely independent of us humans'' (\cite{tegmark:08} p. 101).  If taken literally, however, the ERH violates energy conservation, as it allows human beings to behave arbitrarily without affecting ``external physical reality'' and vice-versa.  The assumption of decompositional equivalence, on the other hand, does not involve, entail, or allow decoupling of observers or any other systems from their environments; any evidence that energy is not conserved, or evidence that energy is conserved but not additive would be evidence that decompositional equivalence is not satisfied in our universe.  Were our universe to fail in fact to satisfy decompositional equivalence, any shift in specified system boundaries - any change in the TPS of $\mathcal{H}_{\mathbf{U}}$ - could be expected to alter fundamental physical laws or their dynamical outcomes; in such a universe, the notions of ``fundamental physical laws'' and ``well-defined dynamics'' would be effectively meaningless.  It is, therefore, assumed in what follows that decompositional equivalence is in fact satisfied in our universe $\mathbf{U}$, and hence that the dynamics $H_{\mathbf{U}}$ is independent of system boundaries.

\subsection{Consequence: System-environment decoherence can have no physical consequences}

The assumption of decompositional equivalence has immediate, but largely unremarked, consequences in two areas: the characterization of system-environment decoherence and the characterization of system identification by observers.  Let us consider decoherence first.  The usual understanding of system-environment decoherence (e.g. \cite{zurek:98, zurek:03, schloss:04, schloss:07}) is that interactions between a system $\mathbf{S}$ and its environment $\mathbf{E}$, where $\mathbf{S \otimes E = U}$ is a TPS of $\mathcal{H}_{\mathbf{U}}$, select eigenstates of the $\mathbf{S}$ - $\mathbf{E}$ interaction $H_{\mathbf{S-E}}$.  Such environmentally-mediated superselection or \textit{einselection} \cite{zurek:81, zurek:82} assures that observations of $\mathbf{S}$ that are mediated by information transfer through $\mathbf{E}$ will reveal eigenstates of $H_{\mathbf{S-E}}$; in the canonical example, observations of macroscopic objects mediated by information transfer through the ambient visible-spectrum photon field reveal eigenstates of position.  From this perspective, it is the quantum mechanism of einselection that underlies the classical notion that the ``environment'' of a system - whether this refers to the ambient environment or to an experimental apparatus - objectively \textit{encodes} the physical state of the system, where ``objectively'' has the sense given in the Ollivier-Poulin-Zurek definition \cite{zurek:04, zurek:05} quoted in \S 2.3.

Two features of this standard account of decoherence deserve emphasis.  First, the idea that the environment einselects particular eigenstates of $\mathbf{S}$ in an observer-independent way - that environmental einselection dependes only on $H_{\mathbf{S-E}}$, where both $\mathbf{S}$ and $\mathbf{E}$ are specified completely independently of observers - allows decoherence to mimic ``collapse'' as a mechanism by which the \textit{world} prepares or creates classical information about particular systems that observers can then detect.  In this picture, as in the traditional Copenhagen picture, observers have nothing to do with what ``systems'' are available to observe: the world - in the decoherence picture, the environment - reveals some systems as ``classical'' and not others.  The sense of ``objectivity'' defined by Ollivier, Poulin and Zurek \cite{zurek:04, zurek:05} depends critically on this assumption; without it, the idea that observers can approach the world ``without prior knowledge'' of the systems it contains becomes uninterpretable.  The second thing to note is that the formal mechanism of ``tracing out the environment'' in decoherence calculations \cite{zurek:98, zurek:03, schloss:04, schloss:07} corresponds physically to an assumption that environmental degrees of freedom are irrelevant to the system-observer interaction, i.e. to an assumption that the physical interaction $H_{\mathbf{S-O}}$, where $\mathbf{O}$ is the observer, is independent of $\mathbf{E}$.  This assumption straightforwardly conflicts with the idea that observation - the $\mathbf{S-O}$ interaction - is \textit{mediated} by $\mathbf{E}$.  This conflict between the formalism of decoherence and its model theory suggests that the trace operation is at best an approximate mathematical representation of the physics of decoherence.

By definition, einselection depends on the Hamiltonian $H_{\mathbf{S-E}}$, which is defined at the boundary, in Hilbert space, between $\mathbf{S}$ and $\mathbf{E}$ \cite{zurek:81, zurek:82}.  In a universe that satisfies decompositional equivalence, this boundary can be shifted arbitrarily without affecting the interactions between quantum degrees of freedom, i.e. without affecting the interaction $H_{ij}$, and hence without affecting the matrix element $\langle i | H_{ij} | j \rangle$, between any pair of degrees of freedom $i$ and $j$ within $\mathbf{U}$.  An arbitrary boundary shift, in other words, has \textit{no physical consequences}.  In particular, a boundary shift that transforms $\mathbf{S \otimes E}$ into an alternative TPS $\mathbf{S}^{\prime} \otimes \mathbf{E}^{\prime}$ has no physical consequences for the values of matrix elements $\langle i | H_{ij} | j \rangle$ where $i$ and $j$ are degrees of freedom within the intersection $\mathbf{E \cap E}^{\prime}$, and hence has no physical consequences for states of $\mathbf{E \cap E}^{\prime}$ or for the classical information that such states encode.  The encodings within $\mathbf{E \cap E}^{\prime}$ of arbitrary states of $\mathbf{S}$ and $\mathbf{S}^{\prime}$, and hence of einselected pointer states of $\mathbf{S}$ and $\mathbf{S}^{\prime}$ are, therefore, entirely independent of the boundaries of these systems, and hence entirely independent of the Hamiltonians $H_{\mathbf{S-E}}$ and $H_{\mathbf{S^{\prime}-E^{\prime}}}$ defined at those boundaries.  The encoding of information about $\mathbf{S}$ in $\mathbf{E}$ is, in other words, entirely a result of the action of $H_{\mathbf{U}} = \mathit{\sum_{ij} H_{ij}}$, and is entirely independent of specified system boundaries or ``emergent'' system-environment interactions definable at such specified boundaries.  

It has been proposed, under the rubric of ``quantum Darwinism'' \cite{zurek:06, zurek:09}, that environmental ``witnessing'' of the pointer states of particular macroscopic systems by einselection explains the observer-independent ``emergence into classicality'' of such systems, and hence explains the observer-independent existence of the ``classical world'' of ordinary human experience (see also \cite{landsman:07, schloss:04, schloss:07}).  In a universe satisfying decompositional equivalence, the einselection of pointer states as eigenstates of system-environment interactions cannot, as shown above, be a \textit{physical} mechanism, and hence cannot underpin an observer-independent ``objective'' \cite{zurek:04, zurek:05} encoding of classical information about some particular systems at the expense of classical information about the states of other possible systems in such a universe.  In a universe satisfying decompositional equivalence, the shared environment encodes the states of all possible embedded systems, or none at all.  The notion that environmental witnessing and quantum Darwinism \textit{explain} the ``emergence of classicality'' collapses in a universe satisfying decompositional equivalence, as both require that einselection physically and observer-independently encode the states of some but not all ``systems'' in the state of $\mathbf{E}$ \cite{fields:12c}.

The physics of continuous fluid flow provides a simple example of decompositional equivalence and its consequences for einselection.  It is commonplace to describe fluid flow in terms of deformable voxels, stipulated to be cubic at some initial time $t_{0}$, that contain some particular collection of molecules.  The stipulation of such a voxel has no effect on the intermolecular interactions between the molecules composing the fluid, whether these molecules are within, outside, or on opposite sides of the boundary of the voxel.  Stipulation of a voxel boundary immediately defines, however, a Hamiltonian $H_{in-out}$ that describes the bulk interaction between the molecules within the voxel and those outside.  This bulk interaction can be viewed as decohering the collective quantum state $|in\rangle$ of the molecules within the voxel, with a decoherence time at room temperature and pressure of substantially less than $10^{-20}$ s \cite{joos-zeh:85}, and as einselecting $|in\rangle$ as an eigenstate of position within the fluid at all subsequent times.  Such einselection prevents the wavefunction $|in(t)\rangle$ from spreading into a macroscopically-extended spatial superposition, just as decoherence and einselection by interplanetary dust, gasses and radiation prevent the wavefunction of Hyperion from doing so \cite{zurek-hyp:06}.  Does the state of the fluid outside the stipulated voxel objectively encode the position of the continuously-deforming voxel boundary at which this einselection takes place?  Could observers with no prior knowledge of the stipulated voxel boundary determine its position by observing the state of the fluid?  Obviously they could not.  

The situation with bulk material objects appears, intuitively, to be different from the fluid-flow situation just described.  When viewed in terms of pairwise-interactions between the quantum degrees of freedom of individual atoms, however, the intuitive difference vanishes.  Consider a uniform sphere of Pb embedded in a solid mass of Plexiglas\textsuperscript{\textregistered} plastic.  The interatomic interactions between Pb, C, O and H atoms are completely independent of whether the Pb sphere, the Pb sphere together with a surrounding spherical shell of plastic, a voxel of Pb entirely within the Pb sphere, or a voxel containing only plastic is considered the ``system of interest.''  The boundary of the system stipulated, in each of these cases, is the site of action of a Hamiltonian $H_{in-out}$ that describes the bulk interaction between the atoms within the stipulated boundary and those outside; the action of this Hamiltonian einselects positional eigenstates of the collective quantum state of the atoms inside the boundary just as it does in the case of a voxel boundary in a fluid.  Observers of the states of some arbitrary sample of the atoms in the plastic part of this combined system would, however, be no more capable of determining the site of a stipulated boundary than observers of some arbitrary sample of the fluid molecules in the previous example.

As a final example, consider observers of the experimental apparatus employed by Brune \textit{et al.} \cite{brune:96} to follow the decoherence of single Rb atoms within an ion trap.  Would an observer unfamiliar with the design or purpose of this apparatus, for example a new graduate student, who observed the behavior of its externally-accessible degrees of freedom - either quantum degrees of freedom or bulk macroscopic degrees of freedom such as pointer positions or readouts from digital displays - be capable of inferring the boundary between the trapped Rb atoms and the apparatus itself, including the magnetic and various electromagnetic fields it generates?   Clearly not.  The boundary between the quantum system comprising the trapped Rb atoms and the quantum system comprising the internal radiative degrees of freedom is stipulated by theory, and this theory must be understood to interpret the behavior of the apparatus as a measurement of decoherence time.  Observers of such an apparatus, in other words, \textit{must} have prior knowledge of the system they are observing and \textit{must} have prior agreements about what the bulk macroscopic states of the system indicate - about what the characters displayed on the readouts mean, for example - to comprehend the operation of the apparatus.  The criteria for ``objectivity'' offered by Ollivier, Poulan and Zurek \cite{zurek:04, zurek:05} and quoted in \S 2.3 above fail utterly in this case, just as they do for the ``objectivity'' of voxel boundaries in fluids or the intuitively ``obvious'' boundary of a Pb sphere embedded in plastic.  As in the previous examples, what counts as the boundary of the ``system of interest'' contained within an ion trap is established by an agreed \textit{convention} among the observers, one that can be changed arbitrarily without changing the \textit{physical dynamics} occurring within the ion trap in any way.

If decoherence has no physical consequences for interaction matrix elements, it can have no consequences for entanglement.  The total entanglement in a quantum universe satisfying decompositional equivalence is, therefore, strictly conserved.  Measurements, in particular, cannot \textit{physically} destroy entanglement, and hence cannot create von Neumann entropy.  The state $| \mathbf{U} \rangle$ can, in this case, be considered to be a pure quantum state with von Neumann entropy of zero at all times.  This situation is in stark contrast to that of a universe in which decompositional equivalence is violated, i.e. a universe in which the dynamics \textit{do} depend on system boundaries, either via a physical process of ``wave-function collapse'' driven by measurement or a physical and therefore ontological ``emergence'' of bounded systems driven by decoherence.  In this latter kind of universe, entanglement is physically destroyed by decoherence and von Neumann entropy objectively increases.  A countervailing physical process that creates entanglement, either between measurements or in regions of weak decoherence, and hence decreases von Neumann entropy must be postulated to prevent such a universe from solidifying into an objectively classical system, a kind of system that our universe demonstrably is not.

\subsection{Consequence: Identification of systems by observers is intrinsically ambiguous}  

While they cannot, without violating decompositional equivalence, physically destroy entanglement, observations nonetheless have real-valued outcomes that can be recorded in classical data structures and reported by one observer to another using classical communication.  If the ``systems'' that these outcome values describe cannot be assumed to be specified for observers by decoherence and environmental witnessing, they must be specified by observers themselves, by the deployment of system-identifying POVMs.  It was argued in \S 2.3 above that both the role of the environment in mediating observations and the \textit{de facto} epistemic position of finite observers support defining system-identifying POVMs not over the particular sets of quantum degrees of freedom - the particular Hilbert spaces and thus TPSs of $\mathbf{U}$ - corresponding to recordable outcome values, but over $\mathbf{U}$ as a whole.  With the assumption of decompositional equivalence, this broad approach to defining POVMs becomes not just advisable but inescapable.  If system boundaries can be shifted arbitrarily without physical consequences, they can be shifted arbitrarily without consequences for the recording of observed outcome values in physical media.  Hence the outcome values recorded following deployment of a POVM must be independent of arbitrary shifts of the boundary within $\mathcal{H}_{\mathbf{U}}$, and hence in the TPS of $\mathbf{U}$, over which the POVM is defined.  This can only be the case if the POVM is not defined over one component of a fixed TPS, but rather over all of $\mathcal{H}_{\mathbf{U}}$.

Recall that any finite observer is restricted to a finite encoding of the outcomes obtained with any POVM; any POVM can be considered a mapping to binary codes of some finite length $N$.  This condition can be met by composing an arbitrary POVM $\lbrace E_{i} \rbrace$ with a nonlinear function $I_{\epsilon}$ such that:

\begin{equation}
\langle \mathbf{U} | \mathit{I}_{\epsilon} E_{\mathit{k}} |\mathbf{U}\rangle = \left\{
   \begin{array}{rl}
   \langle \mathbf{U} | \mathit{E}_{\mathit{k}} |\mathbf{U}\rangle ~\text{if}~
        \langle \mathbf{U} | \mathit{E}_{\mathit{k}} |\mathbf{U}\rangle \geq \epsilon; \\
   0 \qquad ~\text{if}~ \langle \mathbf{U} | \mathit{E}_{\mathit{k}} |\mathbf{U}\rangle < \epsilon
   \end{array} \right. \label{povm}
\end{equation}
for some finite resolution $\epsilon$.  Defining any POVM $\lbrace E_{i} \rbrace$ over all of $\mathcal{H}_{\mathbf{U}}$ as in \eqref{povm} renders the definition of ``system'' implicit: a system $\mathbf{S}$ is \textit{whatever} returns finite outcome values $\alpha^{\mathbf{S}}_{\mathit{i}}$ when acted upon by some POVM $\lbrace E^{\mathbf{S}}_{\mathit{i}} \rbrace$ composed with $I_{\epsilon}$.  The detectable degrees of freedom of such a system are, at some time $t$, the degrees of freedom in the inverse images $Im^{-1} E_{k}^{\mathbf{S}}$ of the components $E_{k}^{\mathbf{S}}$ for which $\langle \mathbf{U} | \mathit{E}_{\mathit{k}}^{\mathbf{S}} |\mathbf{U}\rangle \geq \epsilon$ at $t$.

In general, many TPSs of $\mathcal{H}_{\mathbf{U}}$ will satisfy \eqref{povm} for any given $\lbrace E^{\mathbf{S}}_{\mathit{i}} \rbrace$; the collections of quantum degrees of freedom represented by the ``system'' components of these TPSs will be indistinguishable in principle by an observer deploying $\lbrace E^{\mathbf{S}}_{\mathit{i}} \rbrace$.  Observations in any universe satisfying decompositional equivalence thus satisfy a symmetry, called ``observable-dependent exchange symmetry'' in \cite{fields:12d}: any two systems $\mathbf{S}$ and $\mathbf{T}$ for which a POVM $\lbrace E^{\mathbf{S}}_{\mathit{i}} \rbrace$ returns identical sets of outcome values when composed with $I_{\epsilon}$ can be exchanged arbitrarily without affecting observations carried out using $\lbrace E^{\mathbf{S}}_{\mathit{i}} \rbrace$.  To borrow an example from \cite{fields:12d}, many distinct radioactive sources may appear identical to an observer equipped only with a Geiger counter.  It is shown in \cite{fields:12d} that all observational consequences of the no-cloning theorem, the Kochen-Specker theorem and Bell's theorem follow from observable-dependent exchange symmetry.  Decompositional equivalence is sufficient, therefore, for the universe to \textit{appear} quantum-mechanical, not classical, to finite observers whose means of collecting classical information can be represented by POVMs. 

By imposing observable-dependent exchange symmetry on observers, the assumption of decompositional equivalence removes the final sense in which observational classicality might be regarded as objective classicality: two observers who record the same outcomes can no longer infer that their respective POVMs have detected the same collection of quantum degrees of freedom.  As observable-dependent exchange symmetry applies, in principle, to \textit{all} quantum systems, it applies not just to the ``systems of interest'' to which classically communicated outcome values refer, but to the physical media into which such outcome values are encoded.  The ``measurement problem'' in the current framework is thus the problem of explaining not only how discrete outcome values are obtained from quantum systems, but how classical data structures encoding such values are implemented by the collections of quantum degrees of freedom that constitute communication channels, including the collections of quantum degrees of freedom that constitute the apparently-classical memories of observers.  The measurement problem in this formulation is thus the full problem of understanding LOCC.  This formulation of the measurement problem is similar to those encountered in the multiple worlds \cite{everett:57}, multiple minds \cite{zeh:00} or consistent histories \cite{griffiths:02} formulations of quantum theory, all of which assume purely unitary evolution; however, it rejects the implicit ontological assumption, common to these standard approaches, that ``systems'' and hence TPSs can be regarded as constants across ``branches'' or histories, and therefore rejects the assumption that ``classical communication'' can be taken for granted as being physically unproblematic.

\section{Decoherence as semantics}

\subsection{Decoherence as implemented by a POVM}

If decoherence is not a \textit{physical process} by which the environment creates classical information for observers, what is it?  It is suggested in \cite{fields:12a}, and shown in detail in \cite{fields:12c} that decoherence can be self-consistently and without circularity viewed as a purely informational process, a model-theoretic or \textit{semantic} mapping from quantum states to classical information.  It is, therefore, reasonable to think of decoherence as implemented by a POVM.  To see this, it is useful to reconceptualize observation not as the collection by observers of pre-existing classical information, but as a dynamical outcome of the continuous action by the environment on the physical degrees of freedom composing the observer.  If an arbitrary system $\mathbf{S}$ interacts with its environment $\mathbf{E}$ via a Hamiltonian $H_{\mathbf{S-E}}$, a POVM $\lbrace E^{\mathbf{S}}_{\mathit{i}} \rbrace$ can be defined as a mapping:

\begin{equation}
E^{\mathbf{S}}_{\mathit{k}} : | k \rangle \mapsto \alpha^{\mathbf{S}}_{\mathit{k}} = \sum_{i} \langle i | H_{ik} | k \rangle / \sum_{ij} \langle i | H_{ij} | j \rangle, \label{decoh}
\end{equation} 
where $i$ labels degrees of freedom of $\mathbf{S}$ and $k$ and $j$ label degrees of freedom of $\mathbf{E}$.  This POVM maps each degree of freedom of $\mathbf{E}$ to the real normalized sum of its matrix elements, and hence to its total coupling, with the degrees of freedom of $\mathbf{S}$, and hence naturally represents the encoding of $| \mathbf{S} \rangle$ in $| \mathbf{E} \rangle$.  It thus takes the slogan ``decoherence is continuous measurement by the environment'' literally.

In a universe that satisfies decompositional equivalence, the meanings of ``$\mathbf{S}$'' and ``$\mathbf{E}$'' in \eqref{decoh} can be shifted arbitrarily provided $\mathbf{S \otimes E = U}$.  Suppose an observer $\mathbf{O}$ deploys a POVM $\lbrace E_{i} \rbrace$ defined over $\mathbf{U}$, such that the inverse image $Im^{-1} E_{k}$ is outside $\mathbf{O}$ for all components $E_{k}$ for which $\langle \mathbf{U} | \mathit{E}_{\mathit{k}} |\mathbf{U}\rangle \geq \epsilon$.  In this case, $\mathbf{O}$ can be considered the ``system'' and $\cup_{k} (Im^{-1} E_{k}) \subset \mathbf{U}$ where $\langle \mathbf{U} | \mathit{E}_{\mathit{k}} |\mathbf{U}\rangle \geq \epsilon$ can be considered the ``environment'' in \eqref{decoh}; the Hamiltonian $H_{ik}$ then characterizes the observer-environment interaction, and encodes classical information - the outcome values $\alpha_{k}$ - about $\cup_{k} (Im^{-1} E_{k})$ into $| \mathbf{O} \rangle$.  Hence \eqref{decoh} provides a general definition of decoherence as the deployment of a POVM by an observer.  For observers embedded in a relatively static environment, for which the total observer-environment interaction $\sum_{ik} H_{ik}$ is nearly constant, \eqref{decoh} is reasonably interpreted as defining a single, continuously-deployed POVM.  For observers embedded in highly-variable environments that nonetheless exhibit some periodicity, as most human observers are, it is reasonable to view \eqref{decoh} as describing the deployment of not one but a periodic sequence of POVMs, each normalized over a subset of the environmental degrees of freedom with which $\mathbf{O}$ interacts.  As such a sequence must be finite for a finite observer, a finite observer can only be viewed as decohering his, her or its environment in a finite number of ways.  Hence unlike the ``environment as witness,'' a finite observer as witness can physically encode the states of at most a finite number of distinct ``systems.''  Because the POVMs encoded by finite observers are limited in their resolution by $I_{\epsilon}$, each of the distinct ``systems'' representable by a finite observer is in fact an equivalence class under observable-dependent exchange symmetry.

Using \eqref{decoh}, any collection of Hilbert-subspace boundaries that enclose disjoint collections of degrees of freedom and hence define distinct ``systems'' $\mathbf{S}^{\mu}$ can be represented by a collection of distinct POVMs $\lbrace E^{\mu}_{\mathit{i}} \rbrace$.  The detectable outcome values produced by these POVMs have non-overlapping inverse images; hence they all mutually commute.  If these POVMs are regarded as all acting at each of a sequence of times $t_{i}$, their outcomes at those times can be considered to be a sequence of real vectors $\alpha^{\mu}_{i}$.  These vectors form a consistent decoherent history of the $\mathbf{S}^{\mu}$ at the $t_{i}$, in the sense defined by Griffiths \cite{griffiths:02}.  In a universe in which decoherence is an informational process, the number of such consistent decoherent histories and hence the number of ``classical realms'' \cite{hartle:08} is limited only by the number of distinct sets of subspaces of $\mathcal{H}_{\mathbf{U}}$, i.e. is combinatorial in the number of degrees of freedom of $\mathcal{H}_{\mathbf{U}}$.  Each of these histories, as a discrete time sequence of real vectors, can be regarded as a sequential sample of the state transitions of a classical finite state machine (FSM; \cite{ashby:56}).  As shown by Moore \cite{moore:56}, no finite sequence of observations of an FSM is sufficient to uniquely identify the FSM; hence no finite sample of any decoherent history is sufficient to identify the TPS boundaries at which the POVMs contributing to the history are defined, confirming the observable-dependent exchange symmetry of observations in a universe satisfying decompositional equivalence.

\subsection{Decoherence defines a virtual machine}

A classical \textit{virtual machine} is an abstract machine representable by an algorithm executed on a classical Turing machine \cite{tan:76, hopcroft:79}; any executable item of software, from an operating system to a word processor or a numerical simulation, defines a virtual machine.  An \textit{execution trace} of a virtual machine $\mathbf{V}$ is the sequence of state transitions that $\mathbf{V}$ executes from a some given input state.  Any classical FSM is a classical virtual machine; hence any finite sequence of observations made with a POVM can be represented as an execution trace of a classical virtual machine.  Considering that an arbitrary algorithm $A$ can be employed to choose which of a collection of mutually-commuting POVMs to deploy at a given time point $t_{k}$, it is clear that any consistent decoherent history of $\mathbf{U}$ can be represented as an execution trace of a classical virtual machine.  Hence decoherence can, in general, be represented as a  mapping of $\mathcal{H}_{\mathbf{U}}$ to the space of classical virtual machines, i.e. by a diagram such Fig. 1; as such a mapping takes quantum states to classical information, it can be represented as a POVM $\lbrace E_{i} \rbrace$.  The requirement that this diagram commutes is the requirement that the action of the physical propagator $e^{-( \imath / \hbar )H_{\mathbf{U}} t}$ acting from $t_{n}$ to $t_{n+1}$ is represented, by the mapping $\lbrace E_{i} \rbrace$, as a classical state transition from the $n^{th}$ to the $(n+1)^{th}$ state of some virtual machine $\mathbf{V}$.  This commutativity requirement is fully equivalent to the commutativity requirement that defines consistency of observational histories of $\mathbf{U}$ (e.g. \cite{griffiths:02} Eqn. 10.20).  Hence an evolution $H_{\mathbf{U}}$ is consistent under a decoherence mapping $\lbrace E_{i} \rbrace$ if it can be interpreted as an implementation of a classical virtual machine.

\psset{xunit=1cm,yunit=1cm}
\begin{pspicture}(0,0)(16,7.5)
\put(0.5,6){Virtual machine level:}
\put(0.8,3){Physical state level:}
\put(7,6.5){Virtual Machine $\mathbf{V}$}
\put(5.2,6){...}
\put(6,6.1){\vector(1,0){1}}
\put(7.2,6){$|\mathbf{V}_{\mathit{n}}\rangle$}
\put(8.4,6.1){\vector(1,0){1}}
\put(9.6,6){$|\mathbf{V}_{\mathit{n}+\mathrm{1}}\rangle$}
\put(10.8,6.1){\vector(1,0){1}}
\put(12.2,6){...}

\put(6.3,4.5){$\lbrace E_{i} \rbrace |_{n}$}
\put(7.6,3.5){\vector(0,1){2.2}}
\put(10,3.5){\vector(0,1){2.2}}
\put(10.2,4.5){$\lbrace E_{i}\rbrace |_{n+1} $}

\put(7.9,3.5){$e^{-( \imath / \hbar) H_{U}(t)}$}
\put(5.2,3){...}
\put(6,3.1){\vector(1,0){1}}
\put(7.1,3){$|U(t_{n})\rangle$}
\put(8.4,3.1){\vector(1,0){.8}}
\put(9.3,3){$|U(t_{n+1})\rangle$}
\put(11,3.1){\vector(1,0){.8}}
\put(12.2,3){...}

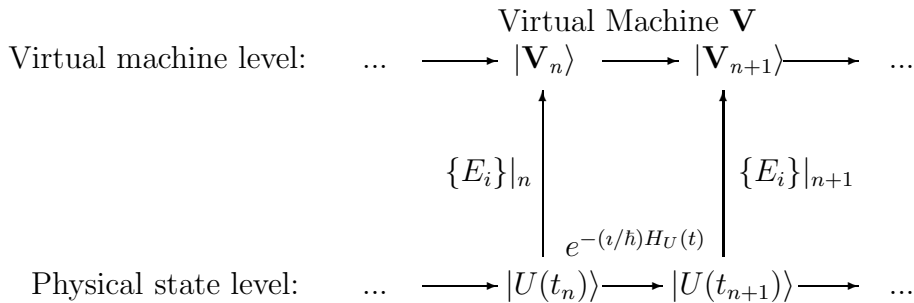
\put(0.5,1.5){\textit{Fig. 1: Semantic relationship between physical states of $\mathbf{U}$ and einselected virtual states $|\mathbf{V}_{\mathit{i}}\rangle$}}
\put(0.5,1){\textit{of a virtual machine $\mathbf{V}$ implemented by $\mathbf{U}$.  Commutativity of this diagram assures that the}}
\put(0.5,0.5){\textit{decoherence mapping $\lbrace E_{i} \rbrace$ is consistent.}}
\end{pspicture}

The semantic relationship shown in Fig. 1 is familiar: it is the relationship by which the behavior of any physical device is interpreted as computation, i.e. as execution of an algorithm characterized as an abstract virtual machine $\mathbf{V}$.  Any consistent decoherence mapping can, therefore, be regarded as an interpretation of the time evolution of $\mathbf{U}$ as classical computation.  As the outcome values returned by any mapping $\lbrace E_{i} \rbrace$ deployed by a finite observer must be collected within a finite time, any such mapping interprets only some \textit{local sample} of the time evolution of $\mathbf{U}$ as computation.  This perspective on decoherence is consistent with the cybernetic intuition - the intuition expressed by the Church-Turing thesis - that any classical dynamical process, and in particular any classical communicative process can be represented algorithmically.

\section{Observation as entanglement}

\subsection{Classical communication is regressive}

We can now return to Alice and Bob, who each perform local observations of a quantum system and then exchange their results by classical communication.  If the dynamics in $\mathbf{U}$ exhibit decompositional equivalence, Alice and Bob cannot rely on decoherence by their shared environment to uniquely identify the system of interest; instead they must each rely on their own POVM to identify it.  Observable-dependent exchange symmetry prevents them, moreover, from determining by observation that they have identified the \textit{same} system of interest; given \eqref{decoh}, they cannot determine without observational access to all degrees of freedom of $\mathbf{U}$ whether they are deploying the same system-identifying POVM.  Under these conditions, what is the meaning of LOCC?

The first thing to note is that any answer to this question that relies on prior agreements between Alice and Bob is straightforwardly regressive, and hence incapable of explaining anything.  How, for example, do Alice and Bob know which POVM to deploy in order to perform a joint observation?  How, in other words, do observers coordinate their observations, independently of whether they manage to observe a single, shared system?  There are two possibilities, as illustrated in Fig. 2.  One involves classical communication: in line with the canonical scenario, some third party presents each observer with a qubit, and instructs them on how to observe it.  The other, more in line with laboratory practice, involves Alice and Bob jointly observing the production of the pair, and then each transporting one of the qubits to a separate site for further observation.  This second option reduces the problem of selecting the correct POVM to employ for the subsequent observations to the problem of resolving the joint system-identification ambiguity when the production of $\mathbf{S}$ is jointly observed.

\psset{xunit=1cm,yunit=1cm}
\begin{pspicture}(0,0)(16,16)
\put(0.5,15){(A)}

\put(1.7,14){Alice}
\put(5.5,14){Bob}
\put(4,11.2){\vector(-1,2){1.3}}
\put(0.5,12.5){``Use $\lbrace A^{\mathbf{S}}_{\mathit{i}} \rbrace$''}
\put(4,11.2){\vector(1,2){1.3}}
\put(5.4,12.5){``Use $\lbrace B^{\mathbf{S}}_{\mathit{i}} \rbrace$''}
\put(3.2,10.7){Classical}
\put(3.4,10.2){Source}
\put(3.8,9.5){$t_{1}$}

\put(9,14){Alice}
\put(14.3,14){Bob}
\put(9.5,11.2){\vector(0,1){2.5}}
\put(8.5,12.5){$\lbrace A^{\mathbf{S}}_{\mathit{i}} \rbrace$}
\put(14.7,11.2){\vector(0,1){2.5}}
\put(14.8,12.5){$\lbrace B^{\mathbf{S}}_{\mathit{i}} \rbrace$}
\put(11.9,10.4){$\mathbf{S}$}

\put(8,10){\line(1,0){8}}
\put(8,10){\line(0,1){1}}
\put(8,11){\line(1,0){8}}
\put(16,10){\line(0,1){1}}

\put(11.9,9.5){$t_{2}$}

\put(4.2,9.6){\vector(1,0){7.5}}
\put(7,9){$e^{-( \imath / \hbar) H_{U}(t)}$}

\put(0.5,8.5){(B)}

\put(1.7,7){Alice}
\put(5.5,7){Bob}
\put(4,4.2){\vector(-1,2){1.3}}
\put(1.7,5.5){$\lbrace A^{\mathbf{S}}_{\mathit{i}} \rbrace$}
\put(4,4.2){\vector(1,2){1.3}}
\put(5.4,5.5){$\lbrace B^{\mathbf{S}}_{\mathit{i}} \rbrace$}

\put(8,3){\line(1,0){8}}
\put(8,3){\line(0,1){1}}
\put(8,4){\line(1,0){8}}
\put(16,3){\line(0,1){1}}
\put(3.9,3.3){$\mathbf{S}$}

\put(3.8,2.5){$t_{1}$}

\put(9,7){Alice}
\put(14.3,7){Bob}
\put(9.5,4.2){\vector(0,1){2.5}}
\put(8.5,5.5){$\lbrace A^{\mathbf{S}}_{\mathit{i}} \rbrace$}
\put(14.7,4.2){\vector(0,1){2.5}}
\put(14.8,5.5){$\lbrace B^{\mathbf{S}}_{\mathit{i}} \rbrace$}
\put(11.9,3.4){$\mathbf{S}$}
\put(11.9,2.5){$t_{2}$}
\put(4.2,2.6){\vector(1,0){7.5}}
\put(7,2){$e^{-( \imath / \hbar) H_{U}(t)}$}

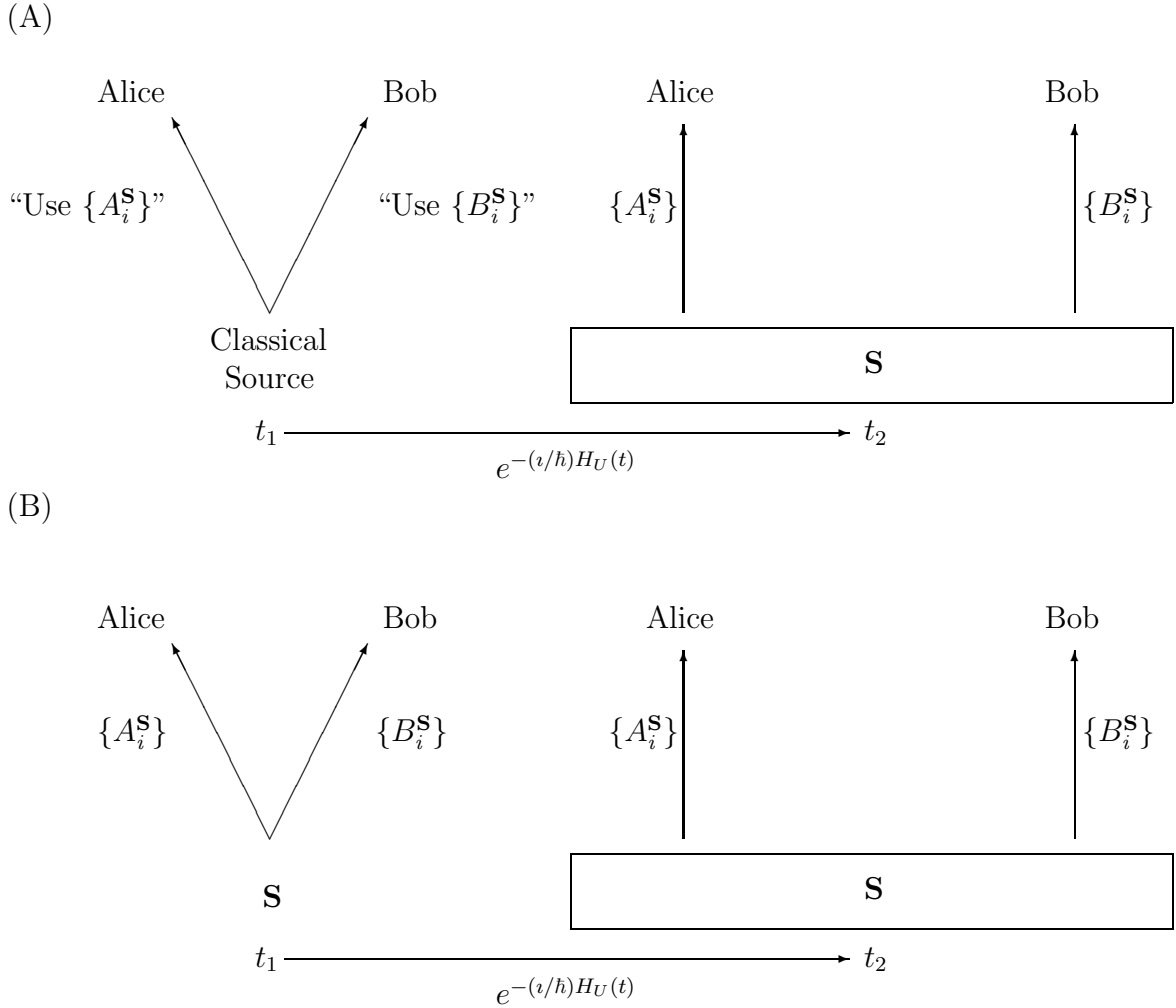
\put(0.2,1.5){\textit{Fig. 2: Two options for coordinating the selection of POVMs $\lbrace A^{\mathbf{S}}_{\mathit{i}} \rbrace$ and $\lbrace B^{\mathbf{S}}_{\mathit{i}} \rbrace$ by Alice and Bob,}}
\put(0.2,1){\textit{respectively.  (A) Alice and Bob receive POVM selection instructions from a classical source.  (B)}}
\put(0.2,0.5){\textit{Alice and Bob jointly observe the production of $\mathbf{S}$ and agree that their selected POVMs identify it.}}
\end{pspicture}

From the perspective of the observers, the two processes illustrated in Fig. 2 both involve the receipt of classical information at $t_{1}$ and its use in directing observations at $t_{2}$; they differ only in the source of the information received at $t_{1}$.  As noted earlier, however, the only means of obtaining classical information provided by quantum theory is the deployment of a POVM.  The two processes differ, therefore, only in \textit{which} POVM the observers deploy at $t_{1}$: in (A) they each deploy a POVM that identifies and determines the state of the ``classical source,'' while in (B) they each deploy a POVM that identifies and determines the state of $\mathbf{S}$.  Hence the coordination question asked at $t_{2}$ can also be asked at $t_{1}$; even if the intrinsic ambiguity of observations with POVMs is ignored, the LOCC scenario cannot get off the ground without an agreement between the observers about which POVM to deploy at $t_{1}$.

In order to reach an agreement about which POVMs to deploy at $t_{1}$, the observers must exchange classical information.  Each observer must, therefore, deploy a POVM that enables the acquisition of classical information from the other; call Alice's POVM for acquiring information from Bob ``$\lbrace A^{\mathbf{B}}_{\mathit{i}} \rbrace$'' and Bob's POVM for acquiring information from Alice ``$\lbrace B^{\mathbf{A}}_{\mathit{i}} \rbrace$,'' and suppose that these POVMs are deployed at some time $t_{0}$.  Clearly the same question can be asked at $t_{0}$ as at $t_{2}$ and $t_{1}$, and clearly it cannot be answered by postulating yet another agreement, another classical communication, and another deployment of POVMs.  The same kind of regress infects any simple joint assumption by Alice and Bob that they are observing the same system, an assumption that must be communicated to be effective.  Any instance of measurement under LOCC conditions, in other words, requires the postulation of \textit{a priori} classical communication between the observers, and hence requires that the observers themselves be regarded as classically objective \textit{a priori}.  Minimal quantum mechanics with decompositional equivalence provides no mechanism by which such \textit{a priori} classical objectivity can be achieved; hence minimal quantum mechanics with decompositional equivalence does not support LOCC.  At best, minimal quantum mechanics with decompositional equivalence supports the \textit{appearance} of LOCC in cases in which observers agree to treat their observations as observations of the same system.

The regress of classical communications encountered here is equivalent to the regress of the von Neumann chain that motivates the adoption of ``collapse'' as a postulate of quantum mechanics \cite{vonNeumann:32}.  Following Everett \cite{everett:57}, the usual response to this regress in the context of minimal quantum mechanics is to postulate observation-induced ``branching'' between the multiple possible outcomes at each instant of observation, with the resulting ``branches'' being regarded as equally ``actual'' either as physically-realized classical universes (e.g. \cite{wallace:10, tegmark:10}) or as classical information-encoding states of a branching observer's consciousness (e.g. \cite{zeh:00}).  In either case, inter-branch decoherence is regarded as conferring observational classicality, and the identity of observed systems across branches is taken for granted; hence decompositional equivalence and observable-dependent exchange symmetry are both violated by the standard Everettian picture.  The concept of branching does not, moreover, explain \textit{how} classical outcomes are encoded by the physical degrees of freedom that implement observers; it therefore leaves open the question of how the communication of classical information is possible. 

\subsection{Memory is communication}

The second thing to note regarding LOCC is that the physical implementation of any classical memory, whether it comprises words written on a page or neural excitation patterns in someone's brain, is a quantum system.  Physically accessing a classical memory requires extracting classical information from this quantum system, and hence requires deploying a POVM.  Observable-dependent exchange symmetry assures that an observer cannot be confidant that the physical degrees of freedom accessed with a ``memory-accessing'' POVM are the same physical degrees of freedom that were accessed when a memory was encoded, or on any previous occasion when the memory was read.  Hence Bob's predicament when accessing his own memory of an observation is no different from Alice's predicament when accessing a report from Bob; in both cases, all the usual caveats pertaining to quantum measurement apply.

The requirement that classical memories be observed in order to function as memories renders the LOCC scenario descriptive of all \textit{reportable} or even \textit{recallable} observations by single observers.  When John Wheeler said ``no phenomenon is a physical phenomenon until it is an observed phenomenon'' (quoted in \cite{scully:07} p. 191), he might as well have said that no phenomenon is a physical phenomenon until it is an observed \textit{and reported} phenomenon, at least reported to the observer him/her/its-self via recall from memory.  It is reporting that renders observational results classical.  In this sense, observational classicality is intrinsically public, or social; without an observer to access a report of an observation, there is no evidence that the observation has been classically recorded.  Hence explaining the appearance of LOCC can be considered to be equivalent to explaining the ability of a single observer to interpret a physical state, including a physical state of his/her/its own memory system, as a classical report of a previous observation.

\subsection{Implementation of POVMs by $H_{\mathbf{U}}$}

Let us suppose that Alice obtains a report from Bob simply by observing his state $|\mathbf{B}\rangle$.  If Alice is to regard a state $| \mathbf{B}_{\mathit{k}} \rangle$ of Bob as a report, i.e. as classically encoding a state $| \mathbf{S}_{\mathit{k}} \rangle$ of some identified external system $\mathbf{S}$, it must be possible, at least in principle, for her to establish that a counterfactual-supporting classical correlation - a classical correlation that exists whether observed or not - between $| \mathbf{S}_{\mathit{k}} \rangle$ and $| \mathbf{B}_{\mathit{k}} \rangle$ is maintained by the $\mathbf{B - S}$ interaction and hence, given decompositional equivalence, by $H_{\mathbf{U}}$.  The action of $H_{\mathbf{U}}$ maintains a counterfactual-supporting classical correlation between states of $\mathbf{S}$ and $\mathbf{B}$ just in case $\mathbf{S}$ and $\mathbf{B}$ are entangled; if the correlation that is maintained is perfect, $\mathbf{S}$ and $\mathbf{B}$ must be monogamously entangled.  Whether joint states of two identified systems appear to be entangled is, however, dependent on the choice of basis and hence the POVM deployed to determine their joint states \cite{zanardi:01, zanardi:04, torre:10, harshman:11, thirring:11}.  Bob's state $| \mathbf{B}_{\mathit{k}} \rangle$ is, therefore, a classical encoding of $| \mathbf{S}_{\mathit{k}} \rangle$ for Alice only if she deploys a POVM that projects $| \mathbf{U} \rangle$ onto a Hilbert-space bases in which $| \mathbf{S} \otimes \mathbf{O} \rangle$ is entangled, and is a perfectly classical encoding if this apparent entanglement is monogamous.

To say of any observer $\mathbf{O}$ that ``$\mathbf{O}$ deploys $\lbrace E^{\mathbf{S}}_{i} \rbrace$ to identify $\mathbf{S}$'' is, therefore, just to say that $\mathbf{O}$ and $\mathbf{S}$ are entangled by the action of $H_{\mathbf{U}}$ on the quantum degrees of freedom that implement $\mathbf{O}$ and $\mathbf{S}$: observation is entanglement.  The existence of such entanglement is an objective fact that is, in a universe satisfying decompositional equivalence, independent of the boundaries of $\mathbf{S}$ and $\mathbf{O}$.  Whether $\mathbf{S}$ and $\mathbf{O}$ appear to be entangled to a third-party observer, however, is not an objective fact; it rather depends on the POVM employed by that observer to extract classical information from the degrees of freedom implementing $\mathbf{S}$ and $\mathbf{O}$.  Hence while the classical correlation between $\mathbf{S}$ and $\mathbf{O}$ is ``real'' - i.e. physical, a result of the action of $H_{\mathbf{U}}$ - whether it appears classical to third parties is virtual, i.e. dependent on semantic interpretation.  All \textit{public} communication is, therefore, nonfungible or ``unspeakable'' in the sense defined in \cite{bartlett:07}: the information communicated is always strictly relative to a POVM - a ``reference frame'' in the language of \cite{bartlett:07} - that is not specified by $H_{\mathbf{U}}$ and cannot be assumed without circularity.  Any publicly-communicable classical description of the world is, therefore, intrinsically logically circular.

The intrinsic circularity of public classical communication renders an explanation of a shared classical world in terms of fundamental physics unattainable.  The shared classical world of ordinary experience cannot, therefore, be regarded as ``emergent'' from fundamental physics alone; instead it must be thought of as \textit{stipulated} by the choice of a POVM, i.e. as stipulated by observers themselves.  From a practical point of view, however, a shared POVM is a shared item of experimental apparatus.  The conclusion that classical communication is entanglement therefore raises the possibility of discovering an item of apparatus that implements a POVM capable of revealing, to third-party observers, the entanglement that transfers classical information from $\mathbf{S}$ to $\mathbf{O}$ in any particular instance.  With such an apparatus, it would be possible to claim a third-party understanding of the local action of $H_{\mathbf{U}}$ that implements any particular instance of classical communication.

\section{Conclusion}

As Bohr \cite{bohr:28} often emphasized, physicists must rely on language, pictures, and other conventionalized tools of human communication to construct descriptions of the world.  They must, moreover, rely on measurements conducted in finite regions of space and time.  The acquisition and communication of classical information is, therefore, always pursued in a LOCC setting.  What has been examined here is the question of how such communication can be understood in terms of basic physics: minimal quantum mechanics together with decompositional equivalence.  It has been shown is that classical communication is quantum entanglement that results deterministically from the action of $H_{\mathbf{U}}$.  Such entanglement is not publicly accessible to multiple observers without the further specification of a POVM.  Any such specification is, however, itself an item of classical information; hence any claim that classical communication ``emerges'' from quantum entanglement involves logical circularity.  The idea that quantum theory can produce a \textit{shared} classicality - can be an ``ultimate theory that needs no modifications to account for the emergence of the classical'' (\cite{zurek:07} p. 1) - therefore cannot be maintained.  This loss of ``emergent classicality'' is, however, balanced by a powerful gain: the possibility that a POVM can be discovered that will reveal, in particular cases, the entanglement by which the transfer of classical information from system to observer is implemented. 

The dependence of physics on model-theoretic or semantic assumptions explored here ties physics explicitly to classical computer science: the selection of a shared POVM that enables quantum theory to get off the ground as a description of a shared observable world is fully equivalent to the selection of a virtual-machine description that enables the description of a physical process as the instantiation of a classical algorithm to get off the ground.  All physical descriptions are, from this point of view, specifications of classical virtual machines.  What distinguishes ``quantum'' from ``classical'' computation is the choice of a POVM.  The increased efficiency of quantum computation is, therefore, not the result of a different kind of device executing a different kind of behavior, but rather the result of a different choice of description.  Castagnoli \cite{castagnoli:10, castagnoli:12} has shown that executions of quantum algorithms can be understood as executions of classical algorithms in which half of the required answer is known up front; what the current analysis suggests is that this half of the required answer is encoded by the POVM with which the initial state of a quantum computation is defined.
 
\bibliographystyle{mdpi}
\makeatletter
\renewcommand\@biblabel[1]{#1. }
\makeatother

\end{document}